\magnification 1200
\centerline {{\bf  Statistical Thermodynamics of Moving Bodies}\footnote*{Based on a lecture  given at 
the Symposium on \lq\lq\ Geometry and Quanta\rq\rq \ ,  held at the University of Torun over the period 
June 25-28, 2008}}  
\vskip 1cm
\centerline {\bf by Geoffrey L. Sewell}
\vskip 0.3cm
\centerline {\bf Department of Physics, Queen Mary University of London,}
\vskip 0.2cm
\centerline {\bf Mile End Road, London E1 4NS: e-mail g.l.sewell@qmul.ac.uk}
\vskip 1cm
\centerline {\it In appreciation of Andrzej Kossakowski\rq s friendship and scientific achievements}
\vskip 0.2cm
\centerline {\it on the occasion of his seventieth birthday }
\vskip 1cm
\centerline {\bf Abstract}
\vskip 0.5cm
We resolve the long standing question of  temperature transformations of uniformly moving bodies by 
means of a quantum statistical treatment centred on the zeroth law of thermodynamics. The key to our 
treatment is the result, established by Kossakowski et al, that a macroscopic body behaves as a thermal 
reservoir with well-defined temperature, in the sense of the zeroth law, if and only if its state satisfies the 
Kubo-Martin-Schwinger (KMS) condition. In order to relate this result to the relativistic thermodynamics 
of moving bodies, we employ the Tomita-Takesaki modular theory to prove that a state cannot satisfy the 
KMS condition with respect to two different inertial frames whose relative velocity is non-zero. This 
implies that the concept of temperature stemming from the zeroth law is restricted to states of bodies in 
their rest frames and thus that there is no law of temperature transformations under Lorentz boosts. The 
corresponding results for nonrelativistic Galilean systems have also been established.
\vfill\eject
\centerline {\bf 1. Introduction}
\vskip 0.3cm
In the wake of Einstein\rq s theory of special relativity, Planck [1] and Einstein [2] proposed an extension 
of classical thermodynamics to bodies moving with uniform velocity $v$ relative to an inertial laboratory 
frame, $K_{L}$. This involved a supplementation of the usual set of thermodynamical variables of a body 
(pressure, volume, temperature, etc.) by this velocity $v$ and led to the result that its temperature $T_{L}$, 
as observed in $K_{L}$, is proportional to the Lorentz contraction factor $(1-v^{2}/c^{2})^{1/2}$; 
specifically that $T_{L}$ is related to the temperature $T_{0}$ of the body relative to  a rest frame 
$K_{0}$ by the formula
$$T_{L}=\bigl(1-v^{2}/c^{2}\bigr)^{1/2}T_{0}.\eqno(1.1)$$  
Evidently, this signifies that a uniformly moving body appears to be cooled by its motion relative to the 
inertial frame of observation.
\vskip 0.2cm
This formula remained unchallenged for more than half a Century until Ott [3] proposed a different 
extension of classical thermodynamics to moving bodies, which led to the opposite result for the 
relationship between $T_{L}$ and $T_{0}$, namely
$$T_{L}=\bigl(1-v^{2}/c^{2}\bigr)^{-1/2}T_{0}.\eqno(1.2)$$  
Subsequently, Landsberg [4] argued, on the basis of another extension of classical thermodynamics, that 
the temperature of the body should be a scalar invariant, i.e. that $T_{L}=T_{0}$. 
\vskip 0.2cm
These different approaches to the problem of extending classical thermodynamics to the relativistic domain 
led to further treatments and comments by a number of authors, e.g. [5-7]. In particular, Van Kampen [5] 
provided a very clear analysis of the underlying assumptions behind the works of [1-4] and proposed yet 
another, relativistically covariant extension of classical thermodynamics. 
\vskip 0.2cm
At this stage we note that all the above works [1-7] were based exclusively on the first and second laws of 
thermodynamics, without reference to either the zeroth law or the underlying statistical mechanics. A 
subsequent work by Landsberg and Matsas [8] invoked both of these latter items in a statistical mechanical 
treatment of a model comprising a two level atom coupled to black body radiation, the atom and radiation 
being at rest in the above described frames $K_{L}$ and $K_{0}$, respectively, and the radiation having a 
Planck spectrum relative to the latter frame. The result they obtained for this model was that the atom is not 
driven to a canonical equilibrium state, relative to $K_{L}$, unless $v=0$. Thus, for this model, the 
temperature concept, as represented by the zeroth law, is applicable only to the radiation in its rest frame. 
This result accords with ideas expressed earlier by Landsberg [4].
\vskip 0.2cm
In this note, as in a previous article [9], we address the question of the generality of this result by means of  
a model independent, quantum statistical treatment of  the response of an arbitrary finite probe 
(thermometer!) $S$, at rest in $K_{L}$, to its coupling to a macroscopic, ideally infinite, system 
${\Sigma}$, which is in thermal equilibrium in its rest frame $K_{0}$. For this setup, we prove that, under 
very general conditions, it is only when ${\Sigma}$ is at rest relative to $K_{L}$ that it drives $S$ into a 
terminal canonical equilibrium state. This signifies that it is only then that ${\Sigma}$ has a well defined 
temperature relative to $K_{L}$, in the sense of the zeroth law. In other words, there is no law of 
temperature transformations under Lorentz boosts. Moreover, a similar argument has established the 
corresponding result for Galilei boosts of nonrelativistic systems [9].
\vskip 0.2cm
The key to these results is the connection, established by Kossakowski, Frigerio, Gorini and Verri [10], 
between the zeroth law of thermodynamics and the Kubo-Martin-Schwinger (KMS )equilibrium condition. 
To explain this connection, we recall that the latter condition on the state of  a conservative quantum 
system, ${\Sigma}$, is given formally be the equation [11]
$${\langle}A(t)B{\rangle}={\langle}BA(t+i{\beta}){\rangle},\eqno(1.3)$$
where ${\langle}.{\rangle}$ denotes expectation value for the state in question, $A$ and $B$ are arbitrary 
observables of the system, $A(t)$ is the evolute of $A$ at time $t$, and ${\beta}$ is the inverse 
temperature in units where ${\hbar}$ and $k_{Boltzmann}$ are equal to unity. The grounds for taking this 
condition to characterise equilibrium states are that
\vskip 0.2cm\noindent
(a) it implies that the state is stationary;
\vskip 0.2cm\noindent
(b) it comprises a generalisation of the canonical Gibbsian condition to infinite systems, which are the 
natural idealisations of macroscopic ones in the standard thermodynamic limit;
\vskip 0.2cm\noindent
(c) it corresponds precisely to various dynamical and thermodynamical stability conditions [12-15] that are 
the natural desiderata for thermal equilibrium; and
\vskip 0.2cm\noindent
(d) it is precisely the condition for which ${\Sigma}$ behaves as a thermal reservoir, in the sense of the 
zeroth law, in that it drives drives any finite test system (thermometer!) $S$ to which it is weakly and 
transitively\footnote*{The transitivity condition is that the ${\Sigma}-S$ coupling induces transitions, 
whether direct or indirect, between all the eigenstates of $S$.} coupled into a terminal state that is the 
canonical equilibrium one of inverse temperature ${\beta}$ [10]. 
\vskip 0.2cm\noindent
It follows from (d) that if a state of ${\Sigma}$ were thermal, in the sense of the zeroth law, from the 
standpoints of observers in both $K_{0}$ and $K_{L}$, then it would satisfy the versions of the KMS 
condition (1.3) relative to both those frames at some inverse temperatures ${\beta}_{0}$ and 
${\beta}_{L}$, respectively. We shall prove, however, that this is not possible, by virtue of the 
mathematical constraints imposed by the KMS condition and the action of Lorentz transformations on the 
observables. Hence we conclude that  there is no law of temperature transformations under Lorentz boosts 
and thus that the concept of temperature stemming from the zeroth law is restricted to states of bodies in 
their rest frames. The corresponding conclusion  for the nonrelativistic setting, with the Lorentz boosts 
replaced by Galilean ones, has also been established [9].
\vskip 0.2cm
We present our treatment of the statistical thermodynamics of moving bodies as follows. In Section 2, we 
formulate the generic operator algebraic model of a relativistic macroscopic system, including a precise 
definition of the KMS condition and its relation to the Tomita-Takesaki modular theory. We then prove, in 
Section 3, that the model cannot support states that satsify the KMS condition relative to two frames of 
reference whose relative velocity is non-zero: this establishes the conclusion described in the previous 
paragraph. In Section 4, we briefly summarise the basis of this conclusion and raise an open question 
concerning the thermodynamics of moving bodies.
\vskip 0.5cm
\centerline {\bf 2. The generic model}
\vskip 0.3cm
We take our model of a relativistic macroscopic system, ${\Sigma}$, to be an infinitely extended one that 
occupies a Minkowski space $X$, whose points we denote by $x$. We formulate the model within the 
operator algebraic framework of Haag and Kastler [16], in which ${\Sigma}$ is represented  by a triple 
$({\cal A},{\cal S},{\alpha})$, where ${\cal A}$ is a $C^{\star}$-algebra of bounded observables, ${\cal 
S}$ is the state space, comprising the positive normalised linear functionals on ${\cal A}$, and ${\alpha}$ 
is a representation of the additive group $X$ (the Minkowski space) in ${\rm Aut}({\cal A})$, 
corresponding to space-time translations. 
\vskip 0.2cm
For a given inertial frame of reference, $K$, we represent the points $x$ of $X$ by coordinates 
${\lbrace}x^{\mu}{\vert}{\mu}=0,1,2,3{\rbrace}$. Here $x^{0}$ is the time coordinate, in units for which 
$c=1$, and the other $x^{\mu}$\rq s are the spatial ones. Thus, the unit vector along the time direction for 
$K$ is 
$$u=(1,0,0,0),\eqno(2.1)$$
and time translations of ${\Sigma}$, relative to $K$, are represented by the one parameter group 
${\lbrace}{\alpha}(tu){\vert}t{\in}{\bf R}{\rbrace}$ of automorphisms of ${\cal A}$. 
\vskip 0.3cm
{\it The KMS Condition and the Modular Automorphisms.} The KMS condition, relative to $K$, on a state 
${\phi}$ may be expressed in the following form [11]. For any $A,B{\in}{\cal A}$, the function 
$F:t({\in}{\bf R}){\rightarrow}{\langle}{\phi};B{\alpha}(tu)A{\rangle}$ extends to the strip 
${\lbrace}z{\in}{\bf C}{\vert}Im(z){\in}[0,{\beta}]{\rbrace}$, where it is analytic in the interior and 
continuous on the boundaries and where
$$F(t+i{\beta})={\langle}{\phi};[{\alpha}(tu)A]B{\rangle} \ {\rm and} \ F(t)= 
{\langle}{\phi};B{\alpha}(tu)A{\rangle} \ {\forall} \ t{\in}{\bf R}.\eqno(2.2)$$
Thus, formally, the KMS condition is simply
$${\langle}{\phi};[{\alpha}(tu)A]B{\rangle}={\langle}{\phi};B{\alpha}((t+i{\beta})u)A{\rangle} \ 
{\forall} \ A,B{\in}{\cal A}, \ t{\in}{\bf R}.\eqno(2.2)^{\prime}$$ 
This condition is closely related to the Tomita-Takesaki theory [17] of modular automorphisms, which 
established that any faithful normal state ${\psi}$ on a $W^{\star}$-algebra ${\cal M}$ induces a unique 
one parameter group, ${\lbrace}{\tau}(t){\vert}t{\in}{\bf R}{\rbrace}$, of automorphisms of ${\cal M}$ 
that satisfies the KMS-like relation
$${\langle}{\psi}[{\tau}(t)M]N{\rangle}={\langle}{\psi}N{\tau}(t+i)N{\rangle} \ {\forall} \ M,N{\in}
{\cal M}, \ t{\in}{\bf R}.\eqno(2.3)$$
\vskip 0.2cm
In order to connect this precisely to the $C^{\star}$-algebraic KMS condition (2.2)$^{\prime}$, we 
introduce the GNS triple $({\cal H},{\pi},{\Phi})$ of the state ${\phi}$ and note that, as this state is 
stationary, the automorphisms ${\alpha}(tu)$ are implemented by the one-parameter group 
${\lbrace}U(t){\vert}t{\in}{\bf R}{\rbrace}$ of unitary transformations of ${\cal H}$ defined by the 
formula [18]
$$U(t){\pi}(A){\Phi}={\pi}({\alpha}(tu)A){\Phi} \ {\forall} \ A{\in}{\cal A}, \ t{\in}{\bf R}.\eqno(2.4)$$
We then define the canonical extensions, ${\tilde {\phi}}$ and ${\tilde {\alpha}}_{u}(t)$, of ${\phi}$ and 
${\alpha}(tu)$, respectively, to ${\pi}({\cal A})^{{\prime}{\prime}}$ by the formulae
$${\tilde {\phi}}(F)=({\Phi},F{\Phi}) \ {\rm and} \ {\tilde {\alpha}}_{u}(t)=U(t)FU(-t) \ 
{\forall} \ F{\in}{\pi}({\cal A})^{{\prime}{\prime}}, \ t{\in}{\bf R}.\eqno(2.5)$$
In particular,
$${\tilde {\alpha}}_{u}(t){\pi}(A)={\pi}({\alpha}(tu)A) \ {\forall} \ A{\in}{\cal A}, \ t{\in}{\bf R}.
\eqno(2.6)$$
It follows from the last two formulae that the KMS condition (2.2)$^{\prime}$ for ${\phi}$ extends to 
${\tilde {\phi}}$ in the form
$${\langle}{\tilde {\phi}};[{\tilde {\alpha}}_{u}(t)F]G{\rangle}=
{\langle}{\tilde {\phi}};F{\tilde {\alpha}}_{u}(t+i{\beta}){\rangle} \ {\forall} \ 
F,G{\in}{\pi}({\cal A})^{{\prime}{\prime}}, \ t{\in}{\bf R}.\eqno(2.7)$$
Moreover, the state ${\tilde {\phi}}$ is faithful [11]. Consequently, it follows from a comparison of Eqs. 
(2.3) and (2.7), with ${\cal M}={\pi}({\cal A})^{{\prime}{\prime}}$ and ${\psi}={\tilde {\phi}}$, that 
the automorphisms ${\tilde {\alpha}}_{u}$ are related to the modulars ${\tau}$ by the formula
$${\tilde {\alpha}}_{u}(t/{\beta})={\tau}(t).\eqno(2.8)$$ 
\vskip 0.5cm
\centerline {\bf 3. Incompatibility of KMS conditions relative to different inertial frames}
\vskip 0.3cm
{\bf Definition 3.1.} We say that space-time translations act non-trivially in a representation ${\pi}$ of 
${\cal A}$ if, for any non-zero $a{\in}X$, there exists a pair $(A,s)$ in ${\cal A}{\times}{\bf R}$ such 
that ${\pi}\bigl({\alpha}(sa)A\bigr){\neq}{\pi}(A)$. 
 \vskip 0.3cm
{\bf Proposition 3.1.} {\it Assume that space-time translations act non-trivially in the GNS representation 
of a state ${\phi}$ on ${\cal A}$. Then ${\phi}$ cannot satisfy KMS conditions with respect to two inertial 
frames whose relative velocity is non-zero.}
\vskip 0.3cm
{\bf Proof.} Let $K$ and $K^{\prime}$ be inertial frames and let $v$ be the velocity of $K^{\prime}$ 
relative to $K$. We choose the spatial coordinate axes so that those of $K^{\prime}$ are parallel to the 
corresponding ones of $K$ and the velocity $v$ is directed along $Ox^{1}$. Then the unit time 
translational vector of $K^{\prime}$, as represented in the $K$ coordinate system, is
$$u^{\prime}=\bigl((1-v^{2})^{-1/2},-v(1-v^{2})^{-1/2},0,0\bigr).\eqno(3.1)$$
\vskip 0.2cm
Suppose now that ${\phi}$ satisfies the KMS conditions relative to both $K$ and $K^{\prime}$ for inverse 
temperatures ${\beta}$ and ${\beta}^{\prime}$, respectively. Then it follows from Eqs. (2.6) and (2.8) that 
both ${\pi}\bigl({\alpha}(tu/{\beta})A\bigr)$ and 
${\pi}\bigl({\alpha}(tu^{\prime}/{\beta}^{\prime})A\bigr))$ are equal to ${\tau}(t){\pi}(A)$. Thus 
$${\pi}\bigl({\alpha}(tu/{\beta})A\bigr)={\pi}\bigl({\alpha}(tu^{\prime}/{\beta}^{\prime})A\bigr) \ 
{\forall} \ A{\in}{\cal A}, \ t{\in}{\bf R}.\eqno(3.2)$$
On replacing $A$ by ${\alpha}(-tu/{\beta})A$ in this formula and invoking the abelian character of the 
space-time translation group, we see that
$${\pi}\Bigl({\alpha}\bigl(t({\beta}{\beta}^{\prime})^{-1}({\beta}u^{\prime}-
{\beta}^{\prime}u)\bigr)A\Bigr)={\pi}(A) \ {\forall} \ A{\in}{\cal A}, \ t{\in}{\cal R}.\eqno(3.3)$$ 
Hence, as space translations are assumed to be non-trivial in the represention ${\pi}$,
$${\beta}u^{\prime}={\beta}^{\prime}u,\eqno(3.4)$$
which, by Eqs. (2.1) and (3.1), implies that
$${\beta}(1-v^{2})^{-1/2}={\beta}^{\prime} \ {\rm and} \ v{\beta}(1-v^{2})^{-1/2}=0.\eqno(3.5)$$
In view of the finiteness of ${\beta}$ and the subluminal condition that ${\vert}v{\vert}<1$, these 
equations cannot be satisfied for non-zero $v$. This completes the proof of the Proposition.
\vskip 0.3cm
{\bf Comments.} (1) Assuming that the laws of thermodynamics are valid in rest frames, it follows from 
this Proposition and Ref. [10] that a state ${\phi}$ that satisfies the zeroth law  relative to these frames does 
not satisfy that law relative to moving ones. Hence the very concept of temperature is restricted to rest 
frames and so there is no law of temperature transformation under Lorentz boosts.
\vskip 0.2cm
(2) The corresponding result for the non-relativistic setting, with Lorentz boosts replaced by Galilean ones, 
has also been established [9] on a similar basis.
\vskip 0.2cm
(3) In view of the commutativity of the space-time translation group, the KMS condition (2.2)$^{\prime}$ 
is equivalent to the formula [19, 20]
$${\langle}{\phi};[{\alpha}(x)A]B{\rangle}={\langle}{\phi};B{\alpha}(x+i{\beta}u)A{\rangle} \ {\forall} 
\ A,B{\in}{\cal A}, \ x{\in}X.\eqno(3.6).$$
This formula may be referred to any inertial frame ${\tilde K}$, with $u$ represented by coordinates 
$({\tilde u}^{0},{\tilde u}^{1},{\tilde u}^{2},{\tilde u}^{3})$. The time component of ${\beta}u$ there is 
then ${\beta}{\tilde u}^{0}$. However, this should {\it not} be taken to be the inverse temperature  relative 
to ${\tilde K}$ if this is not a rest frame, since, by Comment (1), ${\phi}$ does not then satisfy the zeroth 
law for this frame.
\vskip 0.2cm
(4) Since the frames $K$ and $K^{\prime}$ of Prop. 3.1 are both inertial, this Proposition has nothing to 
say about temperatures in accelerating frames or, equivalently, in gravitational fields. Consequently it has 
no bearing on phenomena such as the Hawking and Unruh effects [21-23].  
\vskip 0.5cm
\centerline {\bf 4. Concluding remarks.}
\vskip 0.3cm
We have established that the concept of temperature, which ensues from the zeroth law of thermodynamics, 
is restricted to equilibrium states of  systems in their rest frames. The essential ingredients in the proof of 
this result were the relationships of the KMS condition to the zeroth law of thermodynamics [10] and to the 
Tomita-Takesaki modular theory [19].
\vskip 0.2cm
Granted the validity of classical thermodynamics for systems in their rest frames, the question naturally 
arises whether this discipline may be canonically extended to heterotachic processes comprising exchanges 
of energy and momentum between systems in relative motion. In fact, Van Kampen [5] has initiated an 
approach to this question via a thermodynamical argument to the effect that, at least in certain natural 
model situations, the sum of the entropies of these systems, as defined relative to their rest frames, 
increases in such processes. It would be interesting to have a model independent statistical 
thermodynamical generalisation of this result.
\vskip 0.5cm
\centerline {\bf References.}
\vskip 0.3cm\noindent
[1] M. Planck: Sitzber. K1. Preuss. Akad. Wiss. P. 542, 1907
\vskip 0.2cm\noindent
[2] A. Einstein: Jahrb. Radioaktivitaet Elektronik {\bf 4}, 411, 1907
\vskip 0.2cm\noindent
[3] H. Ott: Zeits. Phys. {\bf 175}, 70, 1963
\vskip 0.2cm\noindent
[4] P. T. Landsberg: Nature {\bf 212}, 571, 1966: Nature {\bf 214}, 903, 1967
\vskip 0.2cm\noindent
[5] N. G. van Kampen: Phys. Rev. {\bf 173}, 295, 1968
\vskip 0.2cm\noindent
[6] T. W. B. Kibble: Nuov. Cim. {\bf 41B}, 72, 1966
\vskip 0.2cm\noindent
[7] H. Kallen and G. Horowitz: Amer. J. Phys. {\bf 39}, 938, 1971
\vskip 0.2cm\noindent
[8] P. T. Landsberg and G. E. A. Matsas: Phys. Lett. A {\bf 223}, 401, 1996
\vskip 0.2cm\noindent
[9] G. L. Sewell: J. Phys. A {\bf 41}, 382003, 2008
\vskip 0.2cm\noindent
[10] A. Kosakowski, A. Frigerio, V. Gorini and M. Verri: Commun. Math. Phys. {\bf 57}, 97, 1977
\vskip 0.2cm\noindent
[11] R. Haag, N. M. Hugenholtz and M. Winnink: Commun. Math. Phys. {\bf 5}, 215, 1967
\vskip 0.2cm\noindent
[12] R. Haag, D. Kastler and E. B. Trych-Pohlmeyer: Commun. Math. Phys. {\bf 56}, 214, 1977 
\vskip 0.2cm\noindent
[13] W. Pusz and S. L. Woronowicz: Commun. Math. Phys. {\bf 58}, 273, 1978
\vskip 0.2cm\noindent
[14] H. Araki and G. L. Sewell: Commun. Math. Phys. {\bf 52}, 103, 1977
\vskip 0.2cm\noindent
[15] G. L. Sewell: Commun. Math. Phys. {\bf 55}, 53, 1977
\vskip 0.2cm\noindent
[16] R. Haag and D. Kastler: J. Math. Phys. {\bf 5}, 848, 1964
\vskip 0.2cm\noindent
[17] M. Takesaki: {\it Tomita\rq s Theory of Modular Hilbert Algebras and its Applications}, Lec. Notes in 
Maths. Vol. 128, Springer, Berlin, Heidelberg, New York, 1970
\vskip 0.2cm\noindent
[18] I. E. Segal: Ann. Math. {\bf 48}, 930, 1947
\vskip 0.2cm\noindent
[19] I. Ojima: Lett. Math. Phys. {\bf 11}, 73, 1986
\vskip 0.2cm\noindent
[20] J. Bros and D. Buchholz: Nucl. Phys. B {\bf 429,}, 291, 1994
\vskip 0.2cm\noindent
[21] S. Hawking: Commun. Math. Phys. {\bf 43}, 199, 1975
\vskip 0.2cm\noindent
[22] W. Unruh: Phys. Rev. D {\bf 14}, 870, 1976 
\vskip 0.2cm\noindent
[23] G. L. Sewell: Ann. Phys. {\bf  141}, 201, 1982
\end